\documentclass[%
 reprint,
 amsmath,amssymb,
 aps,
]{revtex4-2}

\usepackage{graphicx}
\usepackage{dcolumn}
\usepackage{bm}
\usepackage{appendix}
\usepackage{amsmath}

\begin{document}


\title{Evidence of thermodynamics and magnetic monopole plasma formation by photon-magnon interaction in artificial spin ice}

\author{D. G. Duarte}
\affiliation{Laboratory of Spintronics and Nanomagnetism (LabSpiN), Departamento de Física, Universidade Federal de Viçosa, Viçosa-MG, Brazil}

\author{S. F. de Souza}
\affiliation{Laboratory of Spintronics and Nanomagnetism (LabSpiN), Departamento de Física, Universidade Federal de Viçosa, Viçosa-MG, Brazil}

\author{L. B. de Oliveira  }
\affiliation{Laboratory of Spintronics and Nanomagnetism (LabSpiN), Departamento de Física, Universidade Federal de Viçosa, Viçosa-MG, Brazil}

\author{E. B. M. Junior}
\affiliation{Laboratory of Spintronics and Nanomagnetism (LabSpiN), Departamento de Física, Universidade Federal de Viçosa, Viçosa-MG, Brazil}
\author{E. N. D. de Araujo}
\affiliation{Laboratory of Spintronics and Nanomagnetism (LabSpiN), Departamento de Física, Universidade Federal de Viçosa, Viçosa-MG, Brazil}

\author{J. M. Fonseca}
\affiliation{Laboratory of Spintronics and Nanomagnetism (LabSpiN), Departamento de Física, Universidade Federal de Viçosa, Viçosa-MG, Brazil}
\author{C. I. L. de Araujo}%
\email{dearaujo@ufv.br}
\affiliation{Laboratory of Spintronics and Nanomagnetism (LabSpiN), Departamento de Física, Universidade Federal de Viçosa, Viçosa-MG, Brazil}
 
\date{\today}

\begin{abstract}
Artificial spin ices (ASI), containing magnetic monopole quasi-particles emerging at room temperature, have been investigated as a promising system to be applied in alternative low-power information technology devices. However, restrictions associated with the intrinsic energetic connections between opposing magnetic monopoles in conventional ASI need to be overcome to achieve this purpose. Here, photon-magnon scattering in nanomagnets is examined as an approach to locally activate the collective dynamics of interacting magnetic systems at the nanoscale. Low-power white and polarized light were employed as a new tool to manipulate magnetic monopole intensity, leading to tuning on the particles response to external magnetic field and spontaneous magnetization flipping without external field (thermodynamics). Our findings showing evidence of magnetic monopole plasma formation in a regular square ASI system are explained by an analytical model of photon-magnon conversion acting directly on the ASI nanomagnet dipole. Micromagnetic simulations based on the samples parameters and values obtained from the model present a very good qualitative correspondence between theory and observations for the investigated ASI system.

\end{abstract}

\maketitle


\section{Introduction}

Artificial Spin Ice ($ASI$) systems, consisting of nanomagnet arrays nanofabricated in planar geometries in order to present magnetic frustration \cite{wang2006artificial}, have been extensively investigated in the last decade. Such nanofabricated network have the potentiality to mimic, at room temperature \cite{mol2009magnetic, morgan2011thermal}, the properties previously observed only at very low temperatures in natural spin ice Pyrochlore crystals, such as emergency of magnetic monopoles \cite{castelnovo2008magnetic} and thermodynamic phase transitions \cite{bramwell2001spin}.

\begin{figure*}
\centering
\includegraphics[width=13cm]{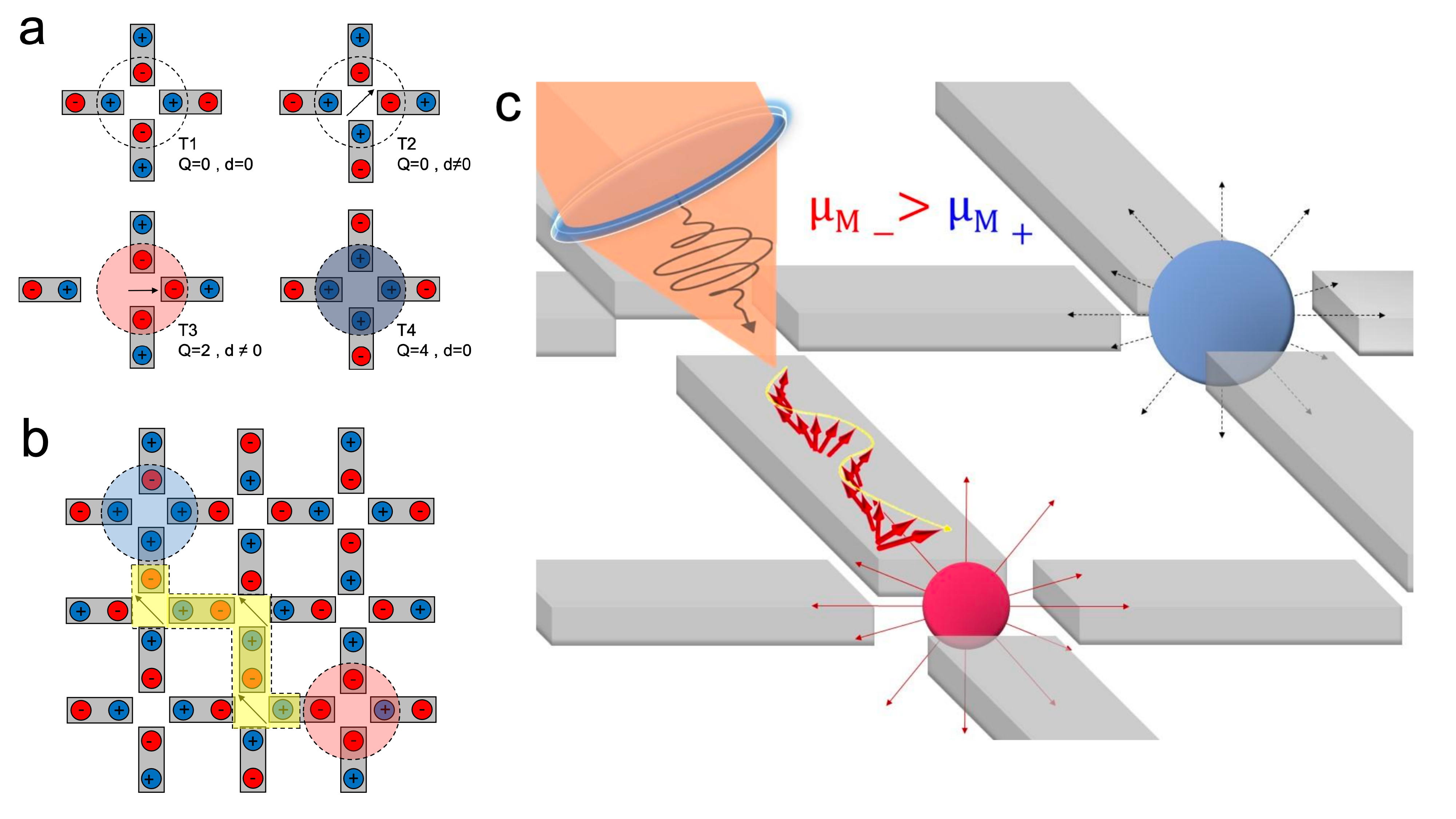}
\caption{{\bf Magnetic monopole mobility control by photon-magnon conversion} {\bf a}, vertex configurations possible in square $ASI$, with ground state in T1 holding zero residual charge, followed energetically by T2 with emergent dipole, T3 configuration presenting both emergent dipole and magnetic charge $q=\pm2$ and T4 with magnetic charge $q=\pm4$. {\bf b}, example of excitation's over the vacuum (ground state), with opposite charges T3 connected by an energetic string composed by T2 configurations and {\bf c} cartoon summering our findings, with a circularly polarized laser generating magnons in the nanoislands, consequently decreasing its dipole moment and magnetic monopole strength.}
\label{fig1}
\end{figure*}

In the most common square $ASI$ system, the four nanomagnets at each vertex can retain magnetic configurations with different energies \cite{wang2006artificial}, as shown in Figure~\ref{fig1}a. By describing the magnetic dipole of each nanomagnet as a magnetic charge dumbbell, it is possible to see that the least energetic configuration T1 has no residual magnetic charge and no magnetic dipole, whereas configuration T2 has no residual magnetic charge but a magnetic dipole at the vertex. Configurations T3 and T4 are the most interesting because they carry residual magnetic charge resembling Nambu magnetic monopoles. Such sort of monopoles have the characteristics of always been attached to an opposite charge by an energetic string composed by T2 configurations to keep the system charge balance \cite{silva2013nambu} (Figure~\ref{fig1}b).  

When it comes to the goal of magnetic monopole information transport in a lower dissipation system, compared to conventional electronics, such an active string may provide a challenge \cite{blundell2012monopoles}. To circumvent this constraint, geometries presenting degeneracy with system vacuum composed by both configurations T1 and T2 have been investigated. These systems allow higher magnetic monopole mobility \cite{goryca2021field, gonccalves2019tuning, loreto2015emergence, morley2019thermally}, because they present a non-energetic string connecting magnetic monopoles \cite{skjaervo2020advances, ribeiro2017realization, loreto2018experimental, perrin2016extensive, farhan2019emergent}, resembling what has been called Dirac monopoles in natural crystals \cite{morris2009dirac}. Some of the major issues that still need to be resolved are the difficulty of building a system ground state manifold with zero magnetic charges at vertices and a controlled method of selectively regulating magnetic charge mobility. 

Recent research has looked into the use of nanomagnets thin enough to be almost superparamagnetic, which allows thermal activation close to room temperature in the context of ground state manifold accomplishment \cite{farhan2013direct}. But the time frame reported to reach the lowest energy state up to this moment is too long to be useful in practical device applications. 

Furthermore, effective methods for manipulating magnetic monopole mobility are currently missing. Recently, we have suggested a method with a reasonable mobility change that involves monopole screening by free electrons on metals \cite{martins2022emergent}. That approach is limited by the permanent charge carriers on metal thin films, resulting in non-reversibility. Regulated methods for quick reversible monopole mobility adjustment in ASI would represent a significant next step in the development of magnetronics. A recent study \cite{pancaldi2019selective} revealed an intriguing method of plasmonic nanomagnets heating by using both constant and pulsed high laser power of 60mW. They showed a reasonable heating and consecutive nanomagnet magnetization change in the timescale of $\mu$s and $ns$ for constant and pulsed light successively. 

In this paper, we will provide a novel approach for tuning magnetic properties in ASI systems using considerably lower laser power, bypassing the nanomagnet heating process. Using an analytical model, we propose that low energetic photon incidence in traditional ASI nanomagnets can produce magnons, which will alter the nanomagnets dipolar energy and hence decrease the monopole strength at the vertex, as illustrated in the cartoon of Figure~\ref{fig1}c. We subsequently demonstrate experimentally, using magnetic force microscopy (MFM) a non-optical magnetic measurement technique, that such an effect can reasonably change the evolution of vertex configurations and system hysteresis as a function of the external magnetic field. Micromagnetic simulations based on materials parameters and estimations from the analytical model were used to support our findings.

Despite of provide a tool for magnetic monopole mobility modification and plasma formation, as it will be further demonstrated, such observed behavior should be considered in general ASI investigations using optical magnetic characterizations, like photoemission electron microscopy combined with magnetic dichroism (PEEM-XMCD) \cite{kapaklis2014thermal, mengotti2008building} and magnetic optic Kerr effect (MOKE) \cite{kohli2011magneto, zhang2013crystallites}, once we show here that light can affect the ASI properties. That could be the reason why predicted ground states were not experimentally fully demonstrated in some ASI systems investigations \cite{gliga2017emergent, loreto2018experimental}.   

\section{Magnon-photon interaction model}
\label{section:mag}

Here, we develop our theoretical model to study photon scattering anomalies in the magnetization of Permalloy nanomagnets. Since we are working with a magnetized medium, prior photon scattering in magnon models is taken into consideration when deriving our Hamiltonian. \cite{ Aspelmeyer2014, Silvia2020, Brahms2010}:

\begin{equation} \label{1}
    \hat{H} = - g \left( \hat{b}^{\dagger}_q + \hat{b}_q \right )\hat{a}^{\dagger}_{\omega} \hat{a}_{\omega^{'}}
\end{equation}
where $g$ is the optomechanical coupling constant, $q = \omega - \omega^{'}$ are magnon and photon frequencies respectively and $\hat{b}$ and $\hat{a}$ are respectively magnon and photon bosonic operators.

Assuming that the photon scatterings causes small disturbances ($\delta \hat{S}<<1$) in the magnetization $\hat{S} = \left(\hat{S}^x, \hat{S}^y, \hat{S}^z \right)$, we can apply the Holstein-Primakoff transformation and obtain $\hat{S}^x = \sqrt{S/2} \left( \hat{b}^{\dagger}_q + \hat{b}_q \right)$, with $S = |\hat{S}|$. Thus, eq (\ref{1}) becomes:
\begin{equation}\label{2}
     \hat{H} = -  G \hat{S}^x \hat{a}^{\dagger}_{\omega}\hat{a}_{\omega^{'}}
\end{equation}
where the optomagnonic coupling constant $G = g\sqrt{2/S}$ is defined \cite{Silvia2020}. Thus, photons scattering into magnons can cause perturbations in the spin; these perturbations are minor deviations in the projection of $\hat{S}$ in the plane $xy$ in the limit $\delta \hat{S} <<1$. Since $G$ depends on the magnetization, it is important to note that the coupling shown in eq.(\ref{2}) only occurs in a medium whose magnetization is non-zero.

Considering an isotropic, non-dissipative medium with a linear magnetization response $\textbf{M}$ \cite{Dey2021}, the permittivity tensor in this case is $\varepsilon_{i,j} \left(\textbf{M} \right) = \varepsilon_0 \left(\varepsilon\delta_{i,j} - if \sum_k \epsilon_{ijk} M_k \right)$, where $\varepsilon_0$ ($\varepsilon$) is the  vacuum (relative) permittivity, $f$ is a material-dependent constant and $\epsilon_{ijk}$ is the Levi-Civita tensor. Using the complex representation of the electric field $\textbf{E} = (\textbf{E}^* + \textbf{E})/2$, the average energy is \cite{Silvia2020}:
\begin{equation}\label{3}
    \Phi = - \dfrac{if\varepsilon_0}{4}\int_V \textbf{M(\textbf{r})}\cdot \left[\textbf{E(\textbf{r})}^*\times \textbf{E(\textbf{r})} \right] d\textbf{r}.
\end{equation}

Light rotates in its polarization when it travels through a magnetic medium; this rotation is related to the permittivity tensor and is determined by the Faraday angle $\theta_F = \omega f M_S / 2c\sqrt{\varepsilon}$, where $c$ denotes the speed of light, $\omega$ denotes the frequency of light, and $M_S$ denotes the saturation magnetization. Conversely, a local effective magnetic field can be produced via the permittivity tensor \cite{Dey2021, Kirilyuk2010}:
\begin{equation}\label{4}
    \textbf{B}_{eff}(\textbf{r},t) = - \dfrac{if\varepsilon_0}{4} \int_V \textbf{E}^*(\textbf{r},t)\times \textbf{E}(\textbf{r}, t) d \textbf{r}.
\end{equation}

This field has a similar impact to spin waves, or magnons. As spin waves are closely related to the light's electric field, we may proceed further by quantize the electric field $\hat{E} = E_{\beta} \hat{a}_{\beta}$, where $E_{\beta}$ denotes the $\beta$ eigenmode of the electric field. A change in the orientation of local magnetization may be caused by the field $B_{eff}$; this phenomenon is known as the Inverse Faraday Effect (IEF)\cite{Dey2021}.

On the other hand, the Holtein-Primakoff representation can be applied if we take into account tiny spin variations. This allows us to directly derive eq(\ref{1}) from eq.(\ref{3}) and, in turn, eq.(\ref{2}), where the optomagnetic coupling is expressed as follows:

\begin{equation}
    G^j_{\beta \gamma} = -i \dfrac{\varepsilon_0 f}{4 \hbar} \sum_{jmn} \int_V \dfrac{M_j(\textbf{r})}{S_j(\textbf{r})} E^*_{\beta m} E_{\gamma n} d\textbf{r}.
\end{equation}

Specifically, we may describe $G$ in terms of the saturation magnetization and the macro spin norm $M_j/M_S = S_j/S$ if we assume that the modes are homogenous (Kittel modes) \cite{Silvia2020} (this is conceivable if the spins are in resonance). The macro spin of the nano-island is represented by the operator $\hat{S}$. $G=c\theta_F \lambda/4S\sqrt{\varepsilon}$, which is the result of decomposing the electric field's eigenmodes on a circular basis \cite{Silvia2020}. For flat waves, lambda is a factor that is close to $1$ \cite{Stancil2009} and $G \backsim 1Hz$ and $S \backsim 10^{10}$ are found in materials like Permalloy \cite{Stancil2009}. The relationship between the total magnon frequency and photon count is given by the constant $G$ \cite{Silvia2020}. Based on this, we can calculate that the maximum magnon density per volume for low power light is around $10^{17} m^{-3}$. A magnon's average wavelength is $1\mu m$. The spin wave magnetic field created by this low power light has an amplitude of around $B_{eff} \backsim 10 mT$, according to equation \ref{2}.

\section{Response to magnetic field in function of white and circularly polarized light}

The studied samples are regular square ASI consisting of Permalloy nanomagnets developed on top of a silicon substrate (see methods \ref{exp}). We validated the reproducibility of our results by performing three sets of measurements in separate areas of $50 \mu m \times 50 \mu m$ for the two specified samples. We will concentrate our analysis on the results from the q04 sample. Equivalent results for sample q08 are presented in Supplementary materials. 
\begin{figure*}
\centering
\includegraphics[width=11cm]{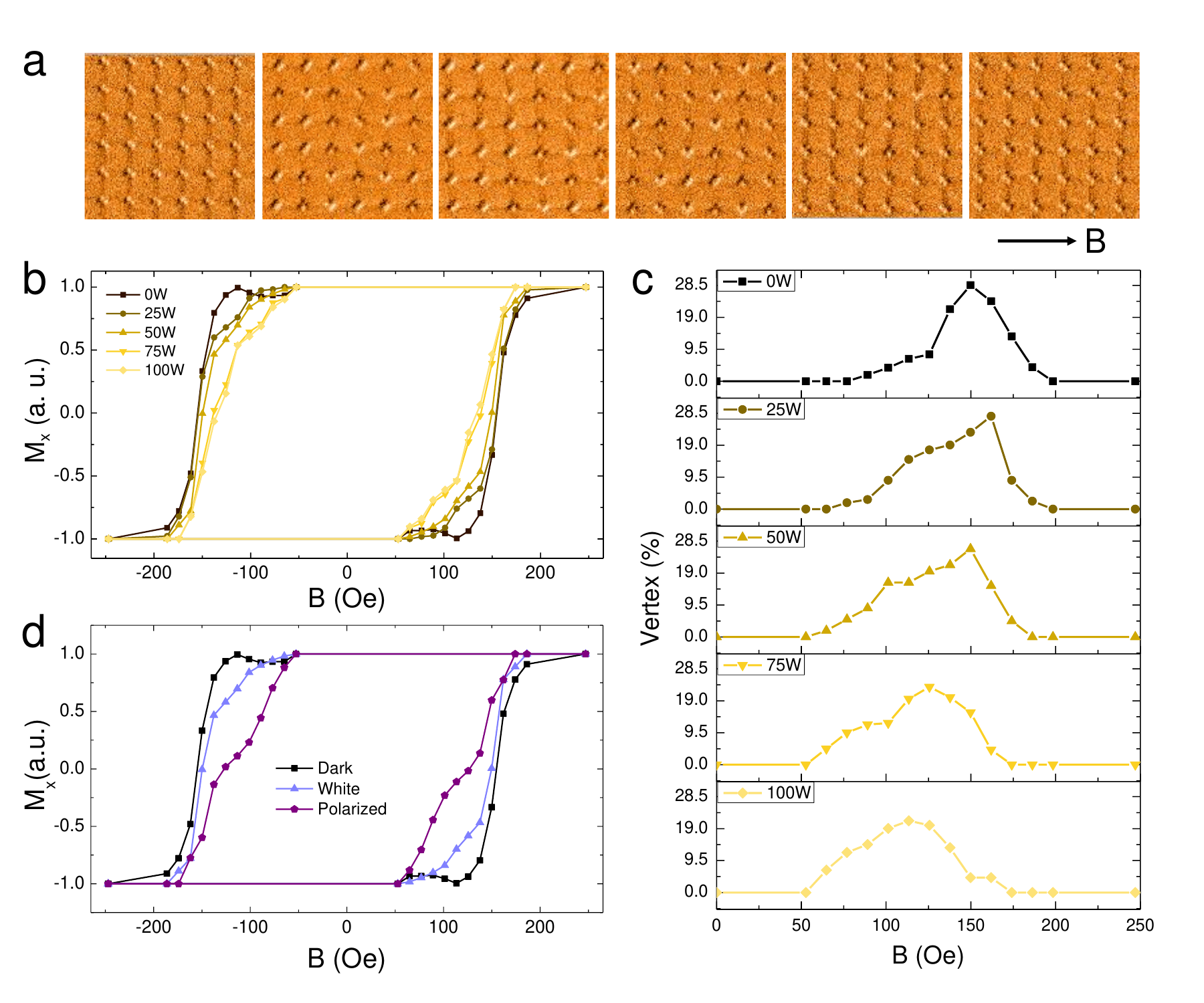}
\caption{{\bf Magnetization reversal and monopole evolution charaterized by MFM measurements} {\bf a}, serie of MFM images obtained during a reversal process performed by external magnetic field applied in the horizontal direction {\bf b}, magnetization reversal in function of white light intensity obtained in remanent state after successive external magnetic field application {\bf c}, change in the magnetic monopoles evolution under different white light power and {\bf d}, comparison between histeretic behavior for the sample on dark and under white and circularly polarized light.}
\label{fig2}
\end{figure*}
Figure~\ref{fig2} shows the characterizations done by magnetic force microscopy (MFM) at room temperature, with the sample in the remanence state after each stage of external magnetic field application. Because the nanomagnet size used here are appropriate for devices exhibiting athermal behavior, the system will maintain its magnetic configuration if not subjected to any external excitation. First, the samples were exposed to an external magnetic field for magnetization saturation in the diagonal direction (top-right to bottom-left). The saturation is confirmed in the first image of blow up MFM images with $20 \mu m \times 20 \mu m$ area given in Figure~\ref{fig2}a.

The stray fields generated by the in plane nanomagnet magnetization results in brilliant and dark patches coming from positive and negative out-of-plane magnetic force on the MFM tip. The other four frames were removed from images obtained during the magnetization reversal process. They were performed with steps of external magnetic field applied in the x-axis direction, in the opposite sense to the initial saturated magnetization. That approach was utilized to study the behavior of individual nanomagnet magnetization flips in relation to the external magnetic field and system internal magnetization, which is primarily determined by geometrical frustration. Two main characteristics are observed in function of external magnetic field: the system's hysterical behavior obtained through the sum of nanomagnet magnetization and magnetic monopoles creation and annihilation obtained by vertex configuration analysis, following the same procedure utilized in our recent works \cite{de2020effects, duarte2022direct, martins2022emergent}. 

We then executed the above-mentioned characterisation in larger areas of $50\mu \times 50\mu$ from dark to a sequence of white light with varying power. The light source was set at a constant distance enough to prevent sample heating during the operation, with constant temperature been measured by a commercial thermocouple. In order to validate the ineffectiveness of light heating in the ASI magnetization we have saturated the nanomagnets magnetization in the sample diagonal, with vertex in the energetic configuration T2, and left it under light exposition for 72 hours. After that period, MFM measurements showed that the imposed saturation state was still there, meaning that the light power was not enough to impose heat and consequently thermodynamics to the system. 

Figure~\ref{fig2}b shows the hysteresis curves obtained by counting each nanomagnet magnetization in x-axys for each magnetic field step applied in the same direction. We illustrate the evolution of nanomagnet magnetization in function of external field under varied white light power. Figure~\ref{fig2}c shows the evolution of magnetic vertex percentage presenting monopole configuration under same conditions. The presented results show a clear influence of depolarized white light power on the magnetic properties of the sample, with a noticeable decrease in the field required to initiate the evolution of magnetic monopoles and a decrease in vertex population density as the light intensity increases. 

To study the effect of light polarization, we reproduced the previous experiment, but this time using 100W of white light with circular polarization. Figure~\ref{fig2}d compares the measurements performed in the dark, white light with 50W of power and circular polarized light with source of 100W, which is cut in half by the polarizer as predicted by the law of Malus.   The large decrease in coercive field observed in the polarized light is compatible with the theoretical prediction presented above for scattering enhancement between photons and magnons. We are going to show latter, by micromagnetic simulations, the influence of generated magnons in the hysteresis process. A lower effect was also studied using light under linear polarization, and a strong direction dependence between polarization and nanomagnet magnetism was found (not shown).  

\section{Thermodynamics produced by focussed polarized light}

As the last stage of the investigation, we will look at possible thermodynamic effects with nanomagnetic magnetization flipping directed solely by light excitation and characteristic ASI internal frustration. In this case, we used a micro Raman spectroscopic setup to excite individual nanomagnets without applying a magnetic field.

\begin{figure*}
\centering
\includegraphics[width=13cm]{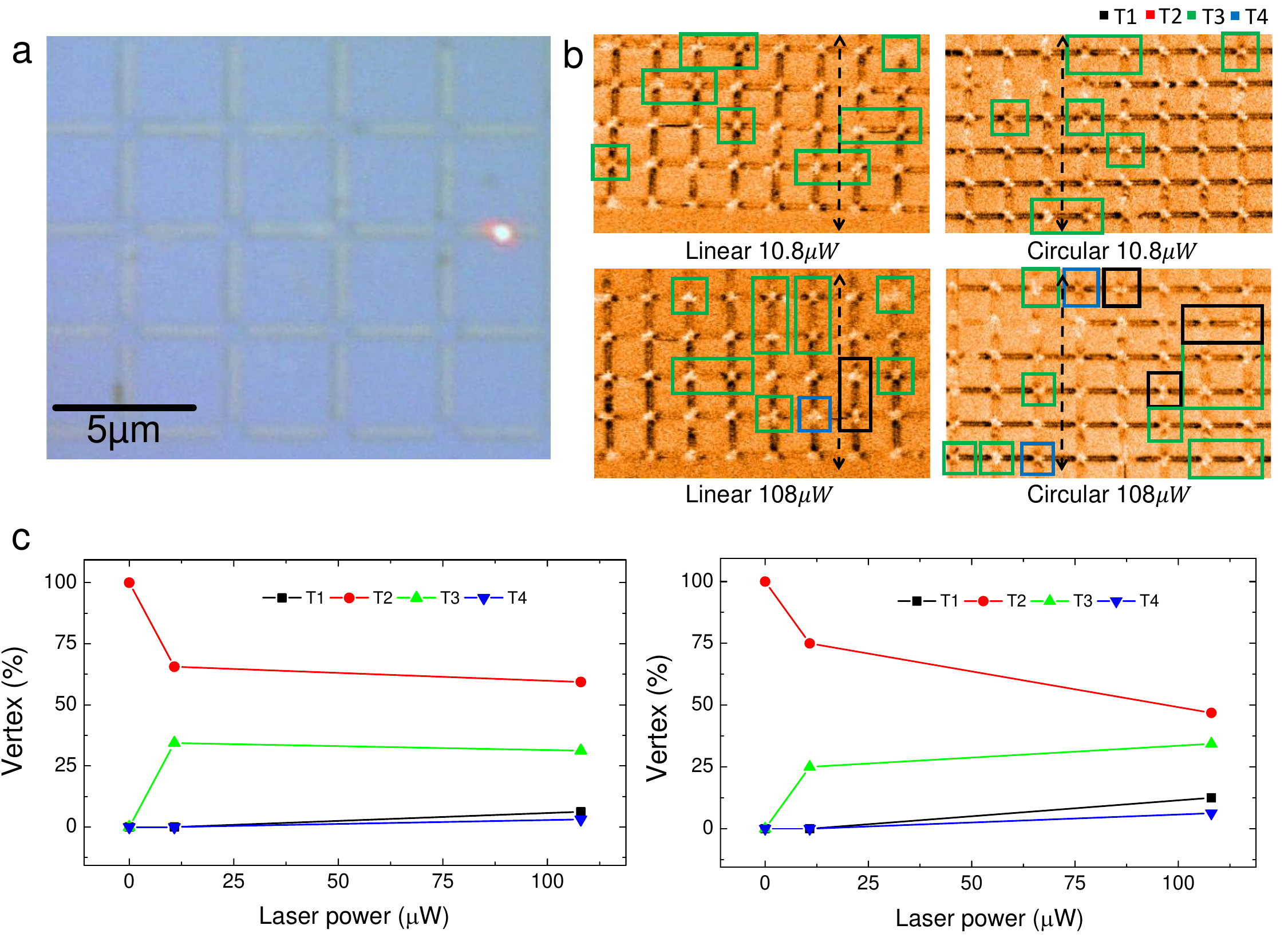}
\caption{{\bf ASI Thermodynamics under linear and circular polarized laser exposition} {\bf a}, Optical image showing the laser spot with average size similar to the nanomagnet width {\bf b}, MFM measurements showing vertex configuration after low power (upper panel) and moderate power (lower panel) for laser exposition with linear (left pannels) and circular (right pannels) polarization. The black dotted lines mark the row of nanomagnets exposed to the laser for 1ns each. {\bf c}, summary of vertex changing after the processes showing thermodynamics effect with much more effectiveness for the circular polarization case on the right panel.}
\label{fig3}
\end{figure*}

In Figure~\ref{fig3}a, we used an argon laser with a wavelength of $632.8nm$ and a beam diameter of $1.03\mu m$ through a $50\times$ objective lens with a numerical aperture of $0.75$. The system's maximum laser power is $2.17mW$, but we are interested in low power regime, so we have utilized for circular polarization both $0.5\%$ and $5\%$ of the utmost capacity, namely $10.85 \mu W$ and $0.108 mW$, with each nanomagnet in a chosen row, marked in Figure~\ref{fig3}a with a red dotted line, subjected to laser exposure for approximately $1$ second. 

Figure~\ref{fig3}b depicts the MFM pictures after laser exposition, as indicated by the red dotted line. The images on the left were acquired after saturated samples exposed to a linearly polarized laser with $10.85 \mu W$ (top) and $0.108 mW$ (bottom), whereas the right panel shows the identical condition with a circular polarized laser. It is easy to see that some vertices altered following low-power laser exposure. At $10.85 \mu W$, higher energy T3 vertices (green square) were seen in both linear and circular polarization. At $0.108 mW$, even higher energetic configurations T4 (blue squares) were achieved in both polarizations. 

Figure~\ref{fig3}c shows the evolution of vertex configurations based on laser power for both linear (top) and circular (bottom) polarizations. In Figure~\ref{fig3}c, it is possible to observe not only the formation of high energetic vertex configurations by moderate laser power, but also the evolution of the vertex density statistics to the relation $T1/T2 \backsim 1/2$, which is expected for degenerated systems suggesting the appearance of Gauge fields with magnetic monopole plasma formation \cite{goddard1978magnetic}. This observation is completely compatible with the underlying theoretical framework described here, which states that increasing power by a factor of ten results in a directly corresponding increase in photon and magnon conversion densities. Magnon creation reduces nanomagnet dipolar interaction, modifying the relation between $J_1$ and $J_2$ in a similar way as it was done before in literature \cite{goryca2021field} using an external magnetic field. 

The tests given in this chapter were rigorously performed three times in different sections of the sample, resulting in data with an error margin of less than 2\%. The optical microscope with focused laser setup allowed for the utilization of light to switch nanoislands due to the laser's increased photon density compared to white light, which lacks coherence in earlier research. The experimental data suggest that at low power non-thermal excitation's mediated by photon assisted magnon generation are responsible for the magnetic switching.

\section{Micromagnetic simulations}

To investigate the influence of magnons generated by photon interaction on nanomagnet magnetization and, as a result, observe the effect of such modification in a $ASI$ system, we performed micro-magnetic simulations using our system architecture and materials parameters, as well as the created magnon intensity recovered from the presented analytical model. The simulation was conducted using an open-access GPU-based micro-magnetic simulation software (Mumax³) \cite{vansteenkiste2011mumax}. 

\begin{figure*}
\centering
\includegraphics[width=13cm]{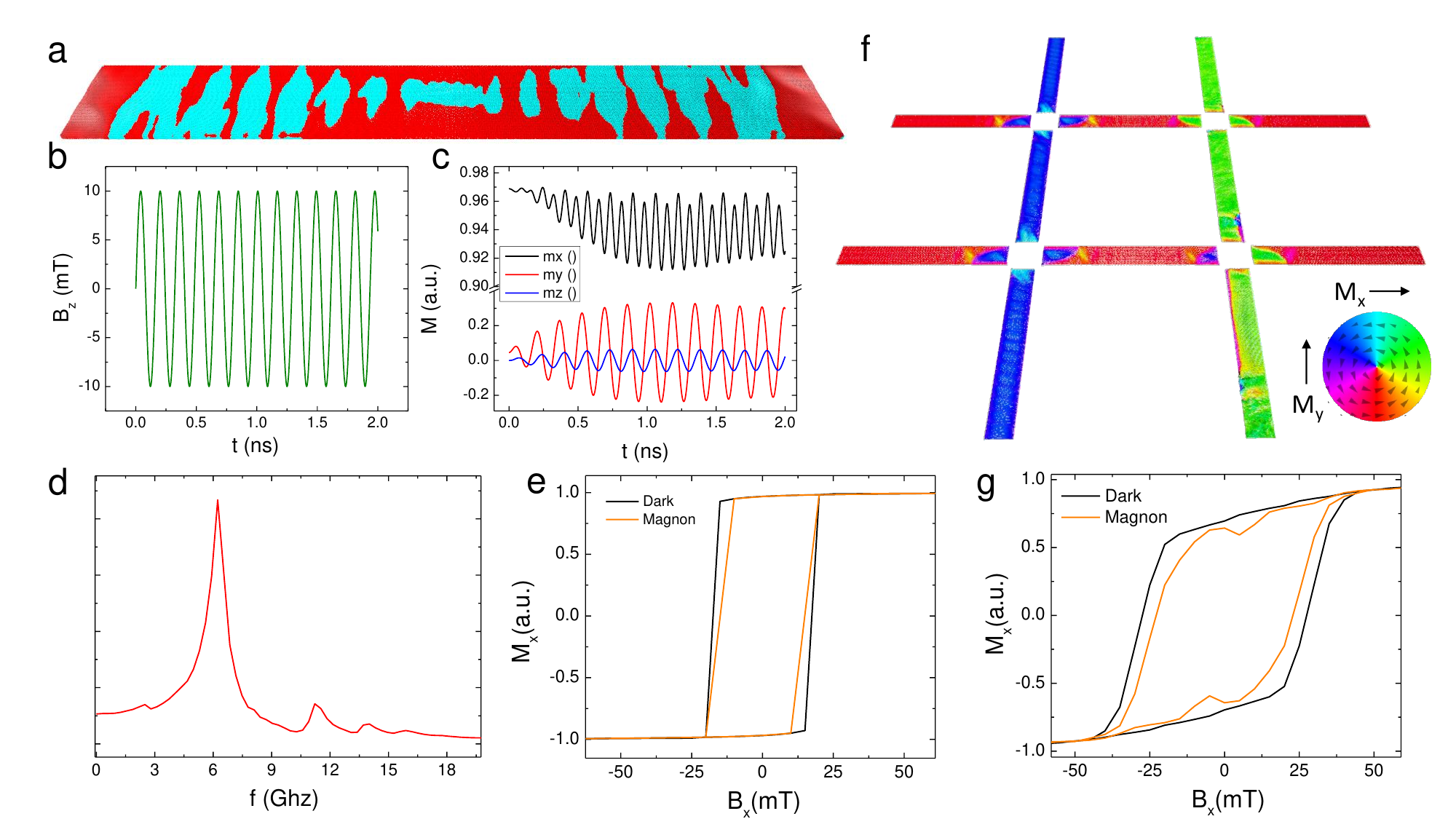}
\caption{{\bf Simulation of magnon influence on nanomagnet magnetization reversal} {\bf a}, Snapshot of magnetization oscillation, with blue and red colors representing out-of-plane component, under {\bf b} ac external magnetic field in z direction. In {\bf c} it is possible to follow the behavior of magnetization in the nanomagnet for x,y and z directions. {\bf d} The frequency of external field excitation for magnon generation was obtained by a protocol of sweep in frequency with magnon modes represented by the peaks. {\bf e} hysteresis sweep with static field applied in x-axys with and without ac external excitation, representing the magnon influence. It is possible to see that the ac excitation decreases the single nanomagnet coercivity. Same protocol was extended for {\bf f}, a portion of an artificial spin ice system under hysteresis sweep with same ac external excitation where the colors represent the in-plane magnetization. {\bf g} simulated hysteresis curve on ASI system presenting similar behavior to the experiments.}
\label{fig4}
\end{figure*}

We first investigate the resonance frequency of a single Permalloy nanoisland used in our studies, which has dimensions of $(3000 \times 400 \times 20)nm$. To do this, we first saturated the magnetization along the nanomagnet's x-axis and then applied a sinusoidal signal in the form $H_{sinc} = \sin{(x)} / x $ perpendicular to the nanomagnet's (z-axis) for $12ns$. Using the described technique, we were able to find the nanomagnet magnon resonance frequency of 6.7 Ghz, as presented in Figure~\ref{fig4}a. This value is in good agreement with values previously presented in literature \cite{gliga2020dynamics, iacocca2020tailoring} and will be used for the simulation of magnon generated on the ASI system. 

As mentioned in section ~\ref{section:mag}, magnons resulting from photon scattering have an associated effective field on the order of $~10mT$. To simulate the effect of magnons on nanomagnet magnetization reversal, we applied an AC external field with the analytically predicted intensity of $~10 mT$ and simulated frequency of 6.7 Ghz on the z-axis direction, while performing a hysteresis cycle with a DC field on the x-axys along the nanomagnet length. Figure~\ref{fig4}b demonstrates that nanomagnet coercivity decreases significantly in the presence of magnons. 

We next applied the same procedure to a tiny ASI lattice with a similar topology as in our experiments. Figure~\ref{fig4}c shows declines in coercivity and magnetization saturation under external field reversal and magnon presence, which are qualitatively similar to the findings given in Figure~\ref{fig4}b.  

\section{Conclusions}

This work describes an analysis of the shine light process in an artificial spin ice system made of normal athermal Permalloy nanomagnets. We observe changes in the ASI histeretical behavior, as well as the evolution of magnetic monopoles, by characterizing magnetization reversal in function of external field under both white and circularly polarized light.

In order to examine possible thermodynamics under light exposure, white light was not enough to provide nanomagnet magnetization flips after 72 hours even in an energetic saturated state. We then exposed each nanomagnet to a low power circularly polarized laser for 1 second, after complete magnetization saturation in the diagonal direction, and observed numerous nanomagnets flip, demonstrating ASI thermodynamics led by light. As a best result, at $0.1 mW$, the new vertex configurations obtained were close to the ratio of the degenerate system, indicating the formation of magnetic monopole plasma under light exposure.

We explain our findings of nanomagnets magnetization switching under the photon excitation using an analytical model that describes magnon creation on nanomagnets via photon conversion. Using micromagnetic simulations, we were able to determine nanomagnet magnon frequency. Finally, utilizing the analytical model's magnon intensity, we simulate the influence of such magnons in the ASI system. The good qualitative match between experimental and theoretical hysteresis curves indicates that our model is successful in explaining the observed shift in monopole mobility due to the action of moderate power light. The same change in the nanomagnet dipole caused by a focussed moderate circularly polarized laser beam may be accountable for the ASI thermodynamics imprinted by light.

\begin{acknowledgments}
The authors thank the Brazilian agencies CNPq 432029/2018-1, FAPEMIG and Coordenação de Aperfeiçoamento de Pessoal de Nível Superior (CAPES)- Finance Code 001 - for the financial support.
\end{acknowledgments}

\bibliography{RefDaniel}

\section{Methods}

\subsection{Sample fabrication}\label{exp}

To prepare the sample, a trilayer of Ta $3 nm$ / Py $20 nm$ / Ta $3 nm$ was sputtered on top of a p-type silicon substrate, with Ta acting as both an adhesion and capping layer. ASI samples in square geometry were manufactured with a total lattice size of $100 \mu m \times 100 \mu m$. The samples q04 and q08 were obtained with different lattice parameters "$a$," which were $800 nm$ and $1200 nm$, respectively. The nanomagnet size of $3000 nm \times 400 nm \times 20 nm$ was carefully created to obtain increased magnetic sensitivity in the Magnetic Force Microscope ($MFM$) up to the limit while ensuring magnetic monodomain in each nanomagnet. 

\subsection{Magnetic force measurements}

The experimental procedure involved initially saturating the nanomagnets' magnetization in the positive $x$-axis direction, followed by stages of magnetic field up to saturation in the opposite direction (negative $x$-axis), without the presence of a light source.
The method was repeated for several levels of white light power: $25\%$, $50\%$, $75\%$, and $100\%$. MFM pictures were consistently captured between each field application in a sample area of $50 \mu m \times 50 \mu m$. To improve experimental dependability, each sample's saturation process was performed three times in various places, with an error margin of 3\%. 

\subsection{Micromagnetic simulations}

Micromagnetic simulation in the square lattices with nine cells was performed with the open-source GPU-based software $MUMAX^3$ 
 \cite{vansteenkiste2011mumax}. The Permalloy numerical parameters utilized in the simulation were magnetic saturation 
 $M_\text{S}=860\times 10^3$ A m$^{-1}$, polarization $P=0.5$, exchange constant $A_\text{ex}=13\times 10^{-12}$ J m$^{-1}$ and Gilbert damping $\alpha=0.01$. Finite difference discretization used for the iterations were based on the Landau--Lifshitz--Gilbert (LLG) equation (Equation~(\ref{eq:LLG}) with cubic cell of $5$ nm $\times$ $5$ nm $\times$ $5$ nm).
\begin{eqnarray}
\frac{\partial\boldsymbol{M}}{\partial t}=\gamma\boldsymbol{H}_\text{eff}\times\boldsymbol{M}
+\frac{\alpha}{M_\text{S}}\boldsymbol{M}\times\frac{\partial\boldsymbol{M}}{\partial t}
-u\frac{\partial\boldsymbol{M}}{\partial y}\nonumber\\+\frac{\beta}{M_\text{S}}\boldsymbol{M}\times\frac{
\partial\boldsymbol{M}}{\partial y}
\label{eq:LLG}
\end{eqnarray}

We investigated magnon resonance in a single nanomagnet using the MUMAXˆ3 catalog technique (vansteenkiste, 2011). We then used our analytical model's intensity and the frequency of the major peak to apply an AC magnetic field out-of-plane in the nanomagnet and ASI system, while completing a hysteretical cycle with a DC field in plane. 

\section{Extended Material}

Here we show the results obtained in the second sample denominated q08. It is possible to see that both behavior under white light are similar to the one presented for sample q04 in Figure~\ref{fig2}.
\begin{figure*}[!hbt]
\centering
\includegraphics[width=12cm]{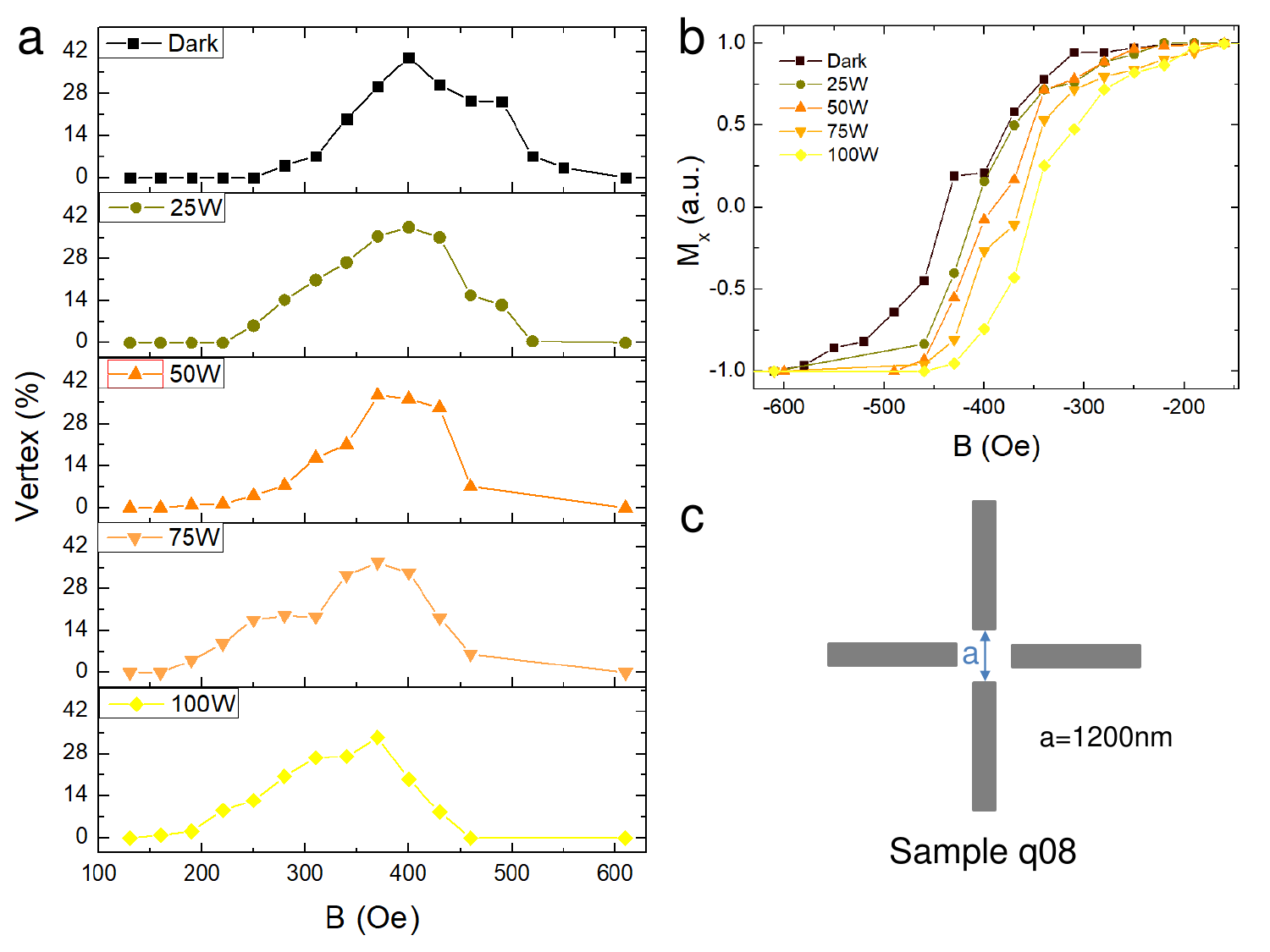}
\label{fig5}
\caption{{\bf Magnetization reversal and monopole evolution charaterized by MFM measurements} {\bf a} change in the magnetic monopoles evolution under different white light power, {\bf b} slope zoom view of magnetization reversal in function of white light intensity obtained in remanent state after successive external magnetic field application  and {\bf c} parameters of the sample q08.}
\end{figure*}

\end{document}